\documentclass[12pt]{revtex4}

\usepackage{epsfig,color,amssymb,hyperref,caption}
\usepackage{float}
\begin{document}
\newcommand{\ltwid}{\mathrel{\raise.3ex\hbox{$<$\kern-.75em\lower1ex\hbox{$\sim$}}}}
\newcommand{\gtwid}{\mathrel{\raise.3ex\hbox{$>$\kern-.75em\lower1ex\hbox{$\sim$}}}}
\newcommand{\bra}{\langle}
\newcommand{\ket}{\rangle}
\newcommand{\sill}{\psi}
\newcommand{\trace}{{\rm Tr}}
\newcommand{\ntilde}{\tilde{n}}
\newcommand{\stilde}{\tilde{s}}
\newcommand{\atilde}{\tilde{\alpha}}
\newcommand{\israel}[1]{\textcolor{red}{#1}}
\newcommand{\adrian}[1]{\textcolor{blue}{#1}}
\def\nn{\nonumber\\}

\bibliographystyle{apsrev}

\title{From frustration to glassiness via quantum fluctuations and random tiling with  exotic entropy}

\author{Israel Klich, Seung-Hun Lee and Kazuki Iida}, 
\affiliation{Department of Physics, University of Virginia, Chalottesville, VA, USA}

\begin{abstract}
\bf{When magnetic moments (spins) are regularly arranged in a geometry of a triangular motif, the spins may not satisfy simultaneously their interactions with their neighbors. This phenomenon, called frustration, leads to numerous energetically equivalent magnetic states (ground states), which results in exotic states such as spin liquid and spin ice. Here we report an alternative situation: a system that, classically, is to be a liquid in the clean limit freezes into a glassy state induced by quantum fluctuations. The case in point is a frustrated magnet in which spins are arranged in a triangular network of bi-pyramids. When taking into account quantum corrections, the classical degeneracy is broken into a set of local minima in a rugged energy landscape, which are separated by large energy barriers, over a finite number of degenerate, periodic, ground states. The appearance of large barriers is due to the absence of local zero-energy modes that are typical in spin-liquid candidate systems. We establish this by mapping the set of local energy minima states into a tiling with colored hexagonal tiles. We show that the system exhibits a large number of aperiodic tessellations. The configuration entropy of the local minima is extremely sensitive to boundary conditions, scaling with the boundary length rather than its volume. The low temperature thermodynamics is also discussed to compare it with other glassy materials. }
\end{abstract}
\maketitle
\newpage
It is well known, since the classical work of Pauling on ice \cite{pauling1935structure}, that certain systems can exhibit an extensive number of energetically equivalent ground states, leading to finite entropy at low temperatures. In a spin ice, states are separated by local single ionic energy barriers, and the spins freeze into one of the equivalent states at low enough temperatures \cite{ramirez1999zero,bramwell2001spin}. In pyrochlore with large spins, locally confined zero energy motions of spins are possible, which can lead to a classical spin liquid state \cite{lee2002emergent}. When quantum effects are taken into account, for small spins, such systems may settle into a super-position of states, forming a quantum spin-liquid\cite{balents2010spin}, as suggested by Anderson \cite{anderson1973resonating}. A closely related but distinct type of systems is glassy systems. One example is amorphous alloys in which the atoms are arranged in a disordered way \cite{schroers2013bulk,greer1993confusion}. Another is spin glass systems in which low concentration of magnetic impurities interact via random long-range interactions \cite{villain1979insulating,mydosh1993spin}. In such systems randomness (or quenching) is the driving force for the freezing phenomena. The randomness, however, makes it difficult to fully understand the complex physics of the freezing phenomenon. Interesting effective models for glassy behavior without disorder have been presented in order to understand various types of glasses. For example, glassy behavior has been explored in systems with long-range interactions \cite{bouchaud1993self} and in models with hard-core classical constraints and stochastic dynamics known as kinetically constrained models\cite{garrahan2010review} (KCM), as well as in certain quantum plaquette and quantum dimer models\cite{castelnovo2005quantum,chamon2005quantum}. Here we show that a spin glass can actually arise from simple nearest neighbor Heisenberg interactions at low temperatures due to quantum effects. Moreover, this behavior may be present in real materials, and may provide a framework to understand the unconventional\cite{lee1996spin} glassy behaviors found in classes of frustrated magnets such as $SrCr_{9p}Ga_{12-9p}O_{19}$ (SCGO(p)),\cite{obradors1988magnetic,ramirez1992elementary,lee1996isolated} and qs-ferrites like $Ba_2Sn_2ZnGa_3Cr_7O_{22}$ (BSZGCO)\cite{hagemann2001geometric} that are highly crystalline and their glassiness seems to be insensitive to disorder\cite{ramirez1992elementary,bono2005correlations}. 

\section{The Model}

Motivated by the seemingly intrinsic nature of the glassy behavior in such crystalline spin glasses, we explore a Heisenberg model on a magnetic lattice realized in SCGO and qs-ferrites. The magnetic lattice of interest is a triangular network of bi-pyramids that are formed by two corner-sharing tetrahedra and are connected by linking triangles (see Fig. 1a, d). Here we consider a simple nearest neighbor spin interaction Hamiltonian $H=\sum_{nn}S_i\cdot S_j$. Classically, any spin configuration in which each tetrahedron and linking triangle has a total zero spin is a ground state. There are an infinite number of energetically equivalent configurations. An important subset of these states is the set of states in which the spins in each bi-pyramid are collinear\cite{iida2012coexisting}. It is well established that collinear configurations are commonly favored in frustrated magnets, which, as we will show later is also the case for the model at hand. Note, in passing, that in most cases co-linearity of spin configurations is global over the entire lattice, while here the collinear direction is not global. Henceforth, we will refer to such states simply as locally collinear (LC) states. 

LC states can be conveniently explored as a simple problem of two degrees of freedom: tri-color (representing the three types of spins for the $120^\circ$ configuration for the AFM linking triangle) and binary sign (representing the parallel (+) or anti parallel (-) direction of each spin within a collinear bi-pyramid of given color (Fig. 1b)). The triangular network of the bi-pyramids forces the tri-color to order long range in a $\sqrt{3}\times\sqrt{3}$ structure as shown by circles in blue, green and red in Fig. 1d. There are 18 possible sign configuration per bi-pyramid, and the sign degrees of freedom are constrained to have the same sign for each linking triangle that connects each three neighboring bi-pyramids \cite{iida2012coexisting}.
\begin{figure*}
\centerline{\includegraphics[width=5in]{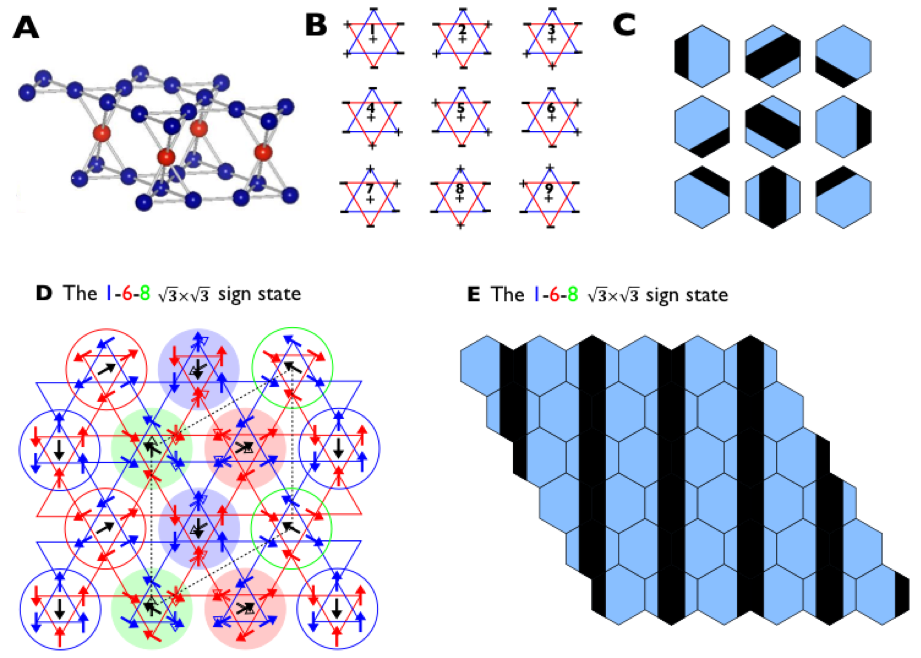}}
\caption{\scriptsize{The triangular network of bi-pyramids and its classical ground states. (A) The lattice is made up by a kagome-triangular-kagome tri-layer. 
(B) In a top view, the tri-layer may be represented as a lattice of bi- pyramids. Each bi-pyramid is represented by two triangles (blue and red), at a 60 degree rotation, corresponding to the upper and lower kagome lattices. The central spin in the middle. In any classical ground state, each tetrahedron has total zero spin. The most important subset of such spin configurations is where spins are individually parallel or antiparallel with the direction specified by the color of the bi-pyramid. These internally collinear states are categorized by assigning for each spin a binary sign (representing a parallel $(+)$ or anti parallel $(-)$ direction relative to the color). These states may be viewed as the 18 elements of $\mathbb{Z}_2\times \mathbb{Z}_3\times \mathbb{Z}_3$ by first specifying the sign of the central spin, and specifying one spin in the upper and the lower layers of the bi-pyramid with the same sign. For simplicity we label the first 9 states numerically 1..9, and their time reversed version as 10..18. (C) A tile representation of sign states. The corners of the hexagon tile represent the spins forming the upper and the lower triangles of the bi-pyramid. The black and blue colors on the boundary represent $(+)$ and $(-)$ signs on the boundary, respectively. (D) A $\sqrt{3}\times\sqrt{3}$ collinear bi-pyramid spin state constructed from the combination of the $\sqrt{3}\times\sqrt{3}$ long range ordered 1-6-8 sign state and a $\sqrt{3}\times\sqrt{3}$ order of the tri-color (the red, blue, green represent the three spin directions of a $120^\circ$ configuration). The unfilled arrows in the middle represent the spins in motion, and the color-shadowed circles represent the affected bi-pyramids. Such spaghetti excitations can act as domain boundaries. (E) The same state as in (D), but in the hexagon representation.}}\label{fig1}
\end{figure*}

\section{Absence of local modes, mapping to a tiling problem and exotic entropy}
Spin liquid candidate systems, such as kagome and pyrochlore, where any spin configurations with zero spin triangles and zero spin tetrahedra, respectively, are ground states, have local zero-energy modes, involving a a finite number of spins, and thus their ground state degeneracy is extensive and scales with the volume. Below, however, we give an entropic argument using a tiling approach to the absence of local-zero modes in our model. Absence of such modes greatly enhances the dynamical barrier to transitions between LC states, and facilitates freezing. 

We map each sign state into a hexagon tile as shown in Fig. 1c. The six corners of the hexagon tile represent the six spins forming the upper and the lower triangles of the bi-pyramid. The tiles are chosen to have the exact matching and enumeration properties of the sign representation (Fig. 1b), spins on the boundary are associated with black and blue colors according to their sign. The middle spin of the bi-pyramid does not interact with other bi-pyramids, and thus we are free to choose the color of the center of the hexagon so as to create the simplest patterns that preserve the topology of the network of positive and negative spins on the boundary. 

Even within the subset of the LC states (sign states), there are numerous ways of covering the entire lattice. With the hexagon representation, the problem of counting the number of sign states in the system becomes a tiling problem. Our tiling problem seems new and bears a remote visual resemblance with a two colored piecewise Herringbone tiling. In order to investigate how the degeneracy increases with the size of the system, we first identified numerically all possible sign states with varying the number of column and rows of bi-pyramids, and thus varying the size of the system. As shown in Fig. 2b, for a given column the number of the possible sign states, $N$, increases with the number of rows in a slower rate than exponentially, which indicates that $N$ does not scale with the volume (area in this quasi-two-dimensional case) of the system. Surprisingly, as shown in the inset, $N$ seems to scale with the number of bi-pyramids on the boundary, i.e., the perimeter of the system. This behavior is corroborated by studying the number of states using transfer matrix methods, we find, numerically, that the largest eigenvalue of the transfer matrix corresponding to a strip of k rows, seems to decrease with k, up to 11 rows, involving 77 spins per unit length. This scaling starkly contrasts with the volume scaling of $N$ of the kagome and pyrochlore systems in which local zero energy modes exist. This non-extensive scaling may be viewed as a consequence of the absence of local zero energy modes. Instead, the smallest unit of zero energy modes scales with the linear dimension of the system, as it involves bi-pyramids along a line, as shown in Fig. 3a. 

\begin{figure*}
\centerline{\includegraphics[width=.7\textwidth]{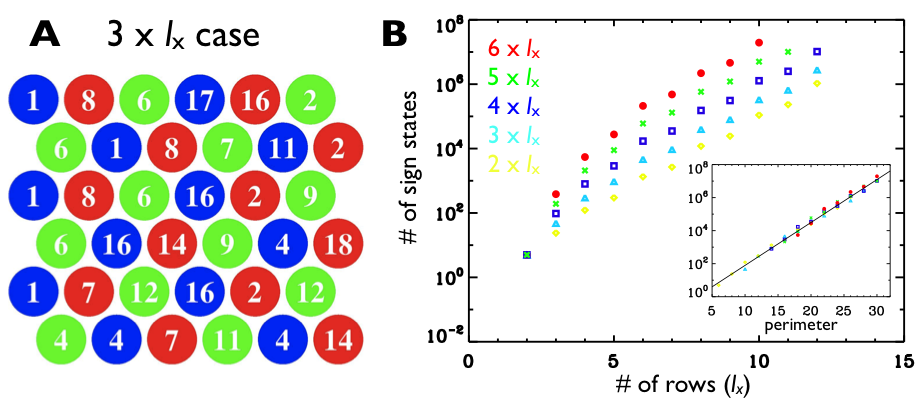}}
\caption{\scriptsize{Counting the possible collinear bipyramid spin states (sign states). (A) One of the possible sign states constructed numerically for the $3\times l_x$ bipyramids where $l_x$ is the number of rows. Each circle with a number represents one bipyramid with the sign state assigned by the number. (B) The number of all possible sign states, $N$, was obtained numerically for different sizes of the system, and plotted in a logarithmic scale as a function of $l_x$. In the inset, $N$ is plotted as a function of the number of bi-pyramids on the boundary.}}\label{fig2}
\end{figure*}

\section{Proving the perimeter scaling of entropy}
The hexagon representation for the sign state of each bi-pyramid allows us to establish bounds on the number of collinear states $N(L,V)$ for a system of volume $V$ and circumference $L$. We find that ${K_1}^L <N(L,V)<{K_2}^LV^{L+1} $, where $K_1, K_2$ are constants. In particular, for $V\sim L^2$, we have: $N(L,V)<{K_2}^L e^{L \log(2L)}$, concluding that (up to a possible logarithmic correction) the number of states is extensive in the boundary length.  

The lower bound is easy to establish: for a given boundary length $L$, we can construct explicitly a number of states which scales as ${K_1}^L$. One way of doing so is by starting from one of the long-range structures as shown in Fig. 1D or 1E. These structures support straight quasi one-dimensional modes that change the state of the bi-pyramid along them. For a square sample of side $L$, we can put up to $L$ independent parallel modes of this type, which supplies us with the lower bound.

To show the upper bound we recast the sign states as a tiling problem. We use the hexagonal tiles depicted in Fig. 1C to obtain a representation of the system as a network of lines. The resultant network may be considered as a fully packed network of rectilinear stripes of alternating color on a lattice, made of straight lines, $\pi/6$ degree turns (``elbows'') and junctions as shown in Fig. 3B. The elements generating this network yield the following properties:  ${\cal P}1$. Lines cannot terminate, and ${\cal P}2$. There are no closed loops. Property ${\cal P}1$ can be verified by inspection of possible termination points, and ruling each of them out. To prove property ${\cal P}2$, assume the contrary and consider a closed loop of black color, inside there must be loops of smaller and smaller sizes. Since the colors alternate, we must have an enclosed simply connected region that is entirely black or entirely blue. Since we do not have an entirely blue or entirely black hexagon in our disposal, such a region must be of limited thickness, therefore the inner region must be made of lines with termination points. By property ${\cal P}1$, such termination points are not allowed.

To proceed, we define a ``laminar region'' as a region where no junctions are present. Property ${\cal P}3$: in a laminar region, by definition, the lines are parallel; moreover, for each of the lines parallel to a chosen reference line (not necessarily a straight one), the thickness at any point along it can be deterministically inferred if the thickness at any other point is known. Property ${\cal P}3$ is established by classifying all possible elbow points that do not involve a junction (Fig. 3B). Properties ${\cal P}1$ and ${\cal P}2$ imply that each line must go through the boundary. Property ${\cal P}3$, shows that in a Laminar region, the thickness degree of freedom of each pattern can be pushed to the boundary, moreover, any elbow must be reflected at two points on boundary of the sample, a detailed study of these properties shows that for a laminar region we can systematically reconstruct the internal state given the boundary of the region. 

Next, we consider junctions to show that $N_J<L$ for any network where $N_J$ is the number of junctions. By properties ${\cal P}1$, ${\cal P}2$, the network is a graph with the only possible termination points on the boundary, and no closed circles: it is thus a forest (disjoint union of trees), with leaves only on the boundary. An elementary fact of graph theory \cite{diestel2005graph} is that the number of nodes in a full binary tree cannot exceed the number of leaves, therefore $N_J<L$. We can have at most $\left(\begin{array}{c} V\\ L \end{array}\right)$ locations for placing junctions in the sample. There are a finite number of possible junction elements. Once the locations and nature of the junctions have been established, the sample excluding the junction is a Laminar region by definition, with effective boundary length proportional to $L+N_J$, following observation ${\cal P}3$, the state is determined by it's boundary. Summing over possible numbers of junctions we have $N(L,V)<\sum_{N_J=0}^L\left(\begin{array}{c} V\\ N_j \end{array}\right) c^{k+N_J }$, which yields the aforementioned upper bound. Thus we have proved that the configurational entropy of collinear states scales with its perimeter rather than its volume.

\begin{figure*}
\centerline{\includegraphics[width=.7\textwidth]{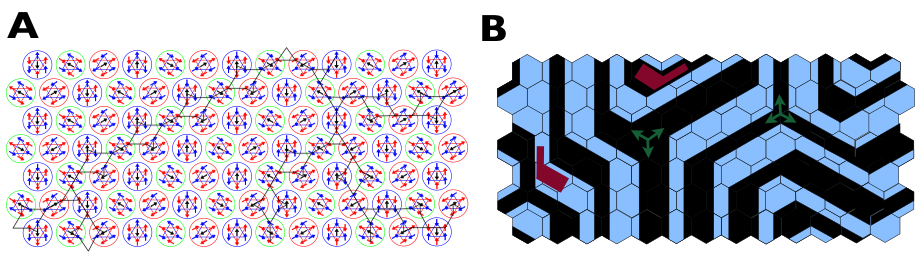}}
\caption{\scriptsize{An arbitrary collinear bipyramid spin state. (A) A collinear bipyramid spin state is shown. At the mean field level, the spins on the black line can rotate collectively without costing any energy, realizing an one-dimensional zero energy mode. (B) The same state as in (A), but in the hexagon representation, that is a network of lines of alternating color. We have marked two elbows by red to show how lines always switch between thin and thick when changing direction, as long as no junction is involved. A couple of junctions are marked by the green arrows.}}\label{fig3}
\end{figure*}

\section{Rugged Energy Landscape and Low Temperatures Thermodynamics}
Let us now turn to the energetics of the sign states. Among the myriad of the sign states, there are six long range ordered states where three types of sign bi-pyramids are arranged in a $\sqrt{3}\times\sqrt{3}$ structure, one of which is the $1-6-8$ state shown in Fig. 1d. Once a sign state is constructed over the entire lattice, the corresponding LC state is constructed by imposing the color ordering. From the LC states, one can generate non-collinear coplanar bi-pyramid states (henceforth coplanar states) by collectively rotating each pair of antiparallel spins in each tetrahedron \cite{iida2012coexisting}. In the mean-field level, the collective motions do not cost any energy, leading to an energy landscape with infinitely large flat bottom formed by collinear and coplanar state and thus to low temperature spin liquid behaviors\cite{sen2012vacancy,arimori2002ordering}.

To investigate what happens when quantum fluctuations are taken into account, we have calculated the energy cost of the quantum fluctuations, within the harmonic (Holstein-Primakoff) approximation around numerous classical spin configurations of minimal energy, with up to $400$ bi-pyramids per sample. This is done by carrying out numerically a symplectic transformation to diagonalize the resultant bosonic Hamiltonians in real space for each state, without assuming long-range order.

An example of the procedure is shown in Fig. 4 for $6 \times 6$ bi-pyramids with several different cdifferent LC states as local minima. Since the long range ordered sign state is special, we considered the LC states near the $\sqrt{3}\times\sqrt{3}$ $1-6-8$ state that are connected with each other through coplanar states. Fig. 4 shows the results; the degeneracy between the collinear and the coplanar states is also lifted, making the LC states local minima and creating energy barriers by the coplanar states. The degeneracy among the LC states is also lifted; the $\sqrt{3}\times\sqrt{3}$ sign state has a lower energy than the other sign states, making the $\sqrt{3}\times\sqrt{3}$ long range ordered state a global minimum and the other LC states local minima. Explicit enumeration shows that there are 6 possible $\sqrt{3}\times\sqrt{3}$ sign states, giving 36 possible $\sqrt{3}\times\sqrt{3}$ spin states when combined with the $6$ possible color configurations. Thus quantum fluctuations lift the mean field ground state degeneracy to form $36$ global minima of the long range ordered LC states and numerous local minima of other LC states, the number of which scales with the perimeter of the system. Since there are no local spin reorientations which connect between the mean field minima, the energy barriers between different states are huge.  As a result, upon cooling, the system gets trapped in one of the local minima of collinear bi-pyramids without a long-range order.

\begin{figure*}
\centerline{\includegraphics[width=.7\textwidth]{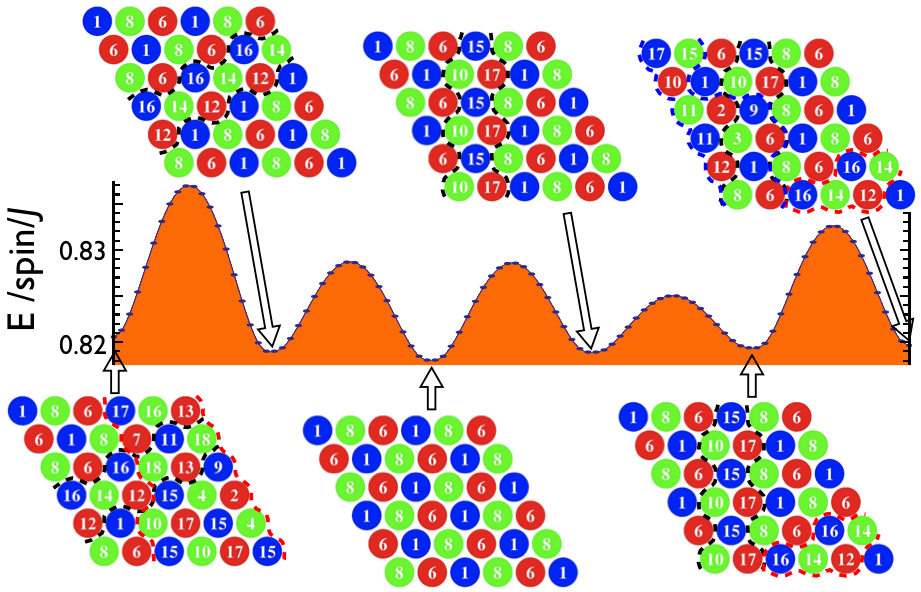}}
\caption{\scriptsize{Rugged energy landscape induced by quantum fluctuations. The magnetic energy of the quantum fluctuations was calculated for several collinear bipyramid spin (sign) states near one global minimum. The energy barriers between the minima are composed of the coplanar bipyramid spin states that connect the sign states.}}\label{fig4}
\end{figure*}

The spin freezing explicitly breaks the $O(3)$ invariance of our Heisenberg Hamiltonian. As a result of this symmetry breaking, and the finite spin stiffness for deforming the aperiodic static antiferromagnetic spin texture, its thermodynamics at low temperatures will be governed by low-energy hydrodynamic Halperin-Saslow modes\cite{halperin1977hydrodynamic}. Such modes are linearly dispersive and lead to a $C_v\propto T^2$ behavior for a quasi-two-dimensional system such as ours. In conventional spin glasses where dilute magnetic ions are embedded in a nonmagnetic metal, there is also a linear in $T$ contribution to the specific heat due to localized two-level systems\cite{anderson1972anomalous}, which dominates its thermodynamics as observed experimentally\cite{mydosh1993spin}. In our system, however, such a linear contribution is negligible at low doping (see also Ref. \cite{podolsky2009halperin}), leading to a $T^2$ behavior at low temperatures, consistent with the experimentally observed behavior in SCGO\cite{ramirez2000entropy}.

\section{Discussion}

The concept of a rugged energy landscape was originally proposed to explain freezing phenomenon found in classic spin glass in which dilute magnetic ions in a nonmagnetic metal interact via long range RKKY interactions that change with distance between the magnetic ions and even change in sign \cite{mydosh1993spin}. The random magnetic interactions induce frustration, which leads to many states of nearly identical energy and a rugged energy landscape. Since then, it has also been suggested to be responsible for other quenching processes that are ubiquitous in nature, ranging from gelation \cite{baumer2013glass} to metallic glass \cite{schroers2013bulk}, to protein folding \cite{bryngelson1987spin}. In such systems, precise mapping of the complex energy landscape as a function of configurations and thus the microscopic mechanism for the freezing phenomena has been challenging. The triangular network of bi-pyramids, on the other hand, does not possess the problem of randomness, and thus provides a unique opportunity to microscopically determine the rugged energy landscape and study the mechanism of the spin freezing, as shown in this work.

An important ingredient in our treatment was the tiling based proof that no classical local zero modes are allowed in the system. We remark that the relevance of tiling as model systems for glassy behavior has been extensively studied for glasses in the context of KCMs\cite{garrahan2010review}. As dynamics is usually allowed only when vacancies in the system are present\cite{Blunt2008Random,Garrahan2009Molecular}, a system that is highly packed (or fully packed as in our discussion here) will be “stuck” in a configuration for a very long time.  Sub-extensive entropy appears in some KCM models. For example the Ising plaquette model on the square lattice exhibits a non-extensive entropy at low temperatures \cite{Jack2005Caging}, In this model glassines is present on the classical level, and $O(3)$ symmetry is broken on the Hamiltonian level. The system is gapped, rendering trivial low temperature thermodynamics. In the context of spin-liquids, a checkerboard model was studied\cite{Tchernyshyov2003Bond}, where in a valance bond solid (VBS) phase, bond configurations are stripe-like, and carry entropy that is extensive in boundary length. However, we note that there, individual spins have no static moment (in fact, spin configurational entropy is extensive in volume in that model). A VBS state was also suggested on a structurally different but related lattice to ours, a \{111\} slice of pyrochlore\cite{Tchernyshyov2004Valence}, which has also a volume scaling entropy. A nematic phase of pseudo-spin was explored in S=1 kagome antiferromagnet with a strong single ion anisotropy \cite{Damle2006Spin,Xu2007Global} which is realized without a static spin moment. 

Finally, we note that similar exotic entropy scaling has been of great interest in other branches of physics, from cosmology, where the entropy of black holes has been argued by Bekenstein and Hawking to scale as the boundary area \cite{bekenstein2003information}, to more recently, in many body quantum mechanical model systems at zero temperature. For example, boundary extensive ground state degeneracy is a feature of some supersymmetric lattice models\cite{huijse2008charge}. A related phenomena is the scaling of entanglement entropy with the boundary area for free scalar fields \cite{bombelli1986quantum}, that obtains logarithmic corrections when a Fermi surface is present \cite{wolf2006violation,gioev2006entanglement}. In our case the perimeter scaling entropy is due to the fact that the local minima of the energy landscape are not separated by local spin rotations (which would typically result in an extensive entropy), but rather are connected with each other by a continuous extended collective rotations of spins, which are sensitive to the states on the boundary. It would be interesting to see if other physical systems possess similar properties.





\begin{acknowledgments}
S.H.L. and I.K. were supported by the Division of Materials Sciences and Engineering, Basic Energy Sciences (BES), US Department of Energy (DE-FG02-10ER46384). I.K. would like to thank Paul Fendley and Assa Auerbach for useful discussions and acknowledges financial support from NSF CAREER award No. DMR-0956053.
\end{acknowledgments}





\newpage 

\renewcommand\thefigure{S\arabic{figure}}

\section{Supplementary Material for: From frustration to glassiness via quantum fluctuations and random
 tiling with exotic entropy}

%


\section{18 possible sign configurations per each collinear bi-pyramid}
The SCGO-like lattice may be viewed as a 2D triangular array of bi-pyramids. This is done by viewing the system top down, as depicted in the Fig. \ref{SCGOLATT} below.

\begin{figure}[H]
\includegraphics[width=4in]{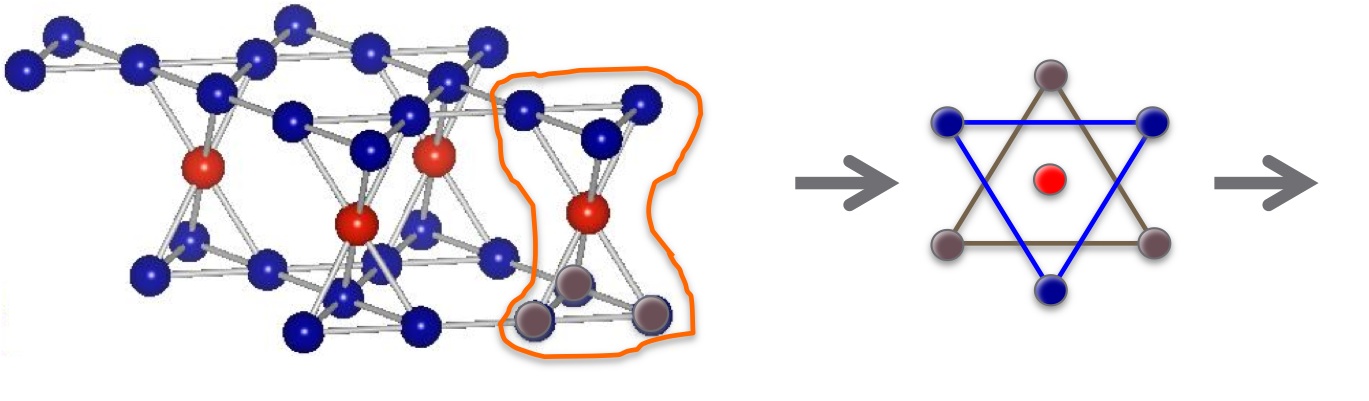}
\includegraphics[width=2in]{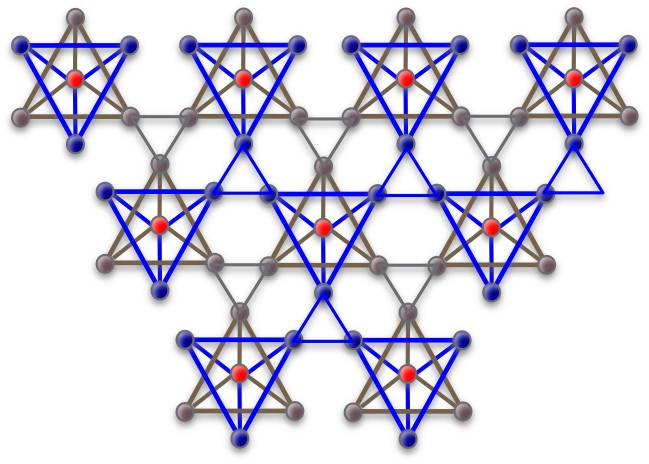}
\caption{The SCGO-like lattice as a triangular array of bi-pyramids.
} \label{SCGOLATT}
\end{figure}
\begin{figure}[H]
\includegraphics*[width=4in]{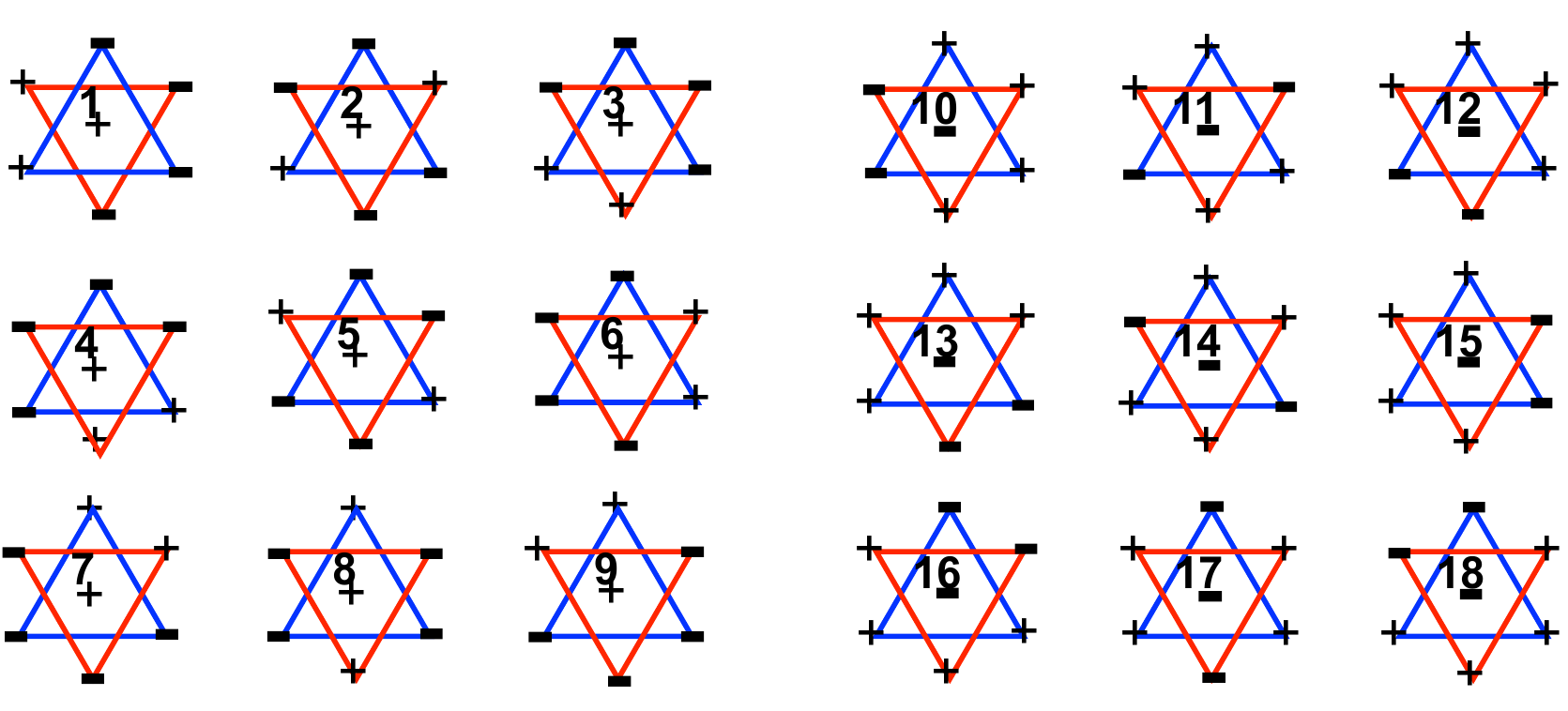}
\caption{18 possible sign configuration per one bi-pyramid. For visualization they are labeled numerically. 
} \label{S1}
\end{figure}

One bi-pyramid consists of two tetrahedrons that share a corner with each other. For a collinear bi-pyramid spin configuration, each spin can be assigned a binary sign, representing a parallel (+) or anti-parallel (-) direction. To satisfy their antiferromagnetic constraints, each collinear tetrahedron must have two plus and two minus, leading to total zero spin. There are 18 possible sign configurations per each bi-pyramid (13), as shown in Fig. \ref{S1}. 

\section{Color-sign or collinear bi-pyramid states}

When the $\sqrt{3}\times\sqrt{3}$ color order is imposed on a sign state, a collinear bi-pyramid state is obtained. Fig. \ref{S2}A illustrates an example with a sign state that does not have any long range order, yielding a collinear bi-pyramid state without any long range order. 

\begin{figure}[H]
\includegraphics*[width=3in]{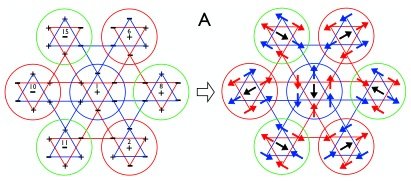}~~~~~
\includegraphics*[width=3in]{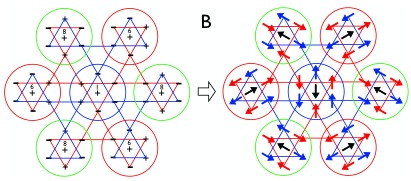}
\caption{The way to obtain a collinear bi-pyramid state from a color-sign state. (A) A collinear bi-pyramid state from a random sign state. (B) A $\sqrt{3}\times\sqrt{3}$ long range ordered collinear bi-pyramid state from the 1-6-8 sign state.
} \label{S2}
\end{figure}
Fig.  \ref{S2}B,  \ref{S2cont}C, and  \ref{S2cont}D show three cases with $\sqrt{3}\times\sqrt{3}$ long range ordered sign states that yield three $\sqrt{3}\times\sqrt{3}$ long range ordered collinear bi-pyramid states. Sign flip operation on the three long range ordered states yields another set of three $\sqrt{3}\times\sqrt{3}$ collinear bi-pyramid states of the 10-15-17, 11-13-18, and 12-14-16 sign states.
\begin{figure}[H]
\includegraphics*[width=3in]{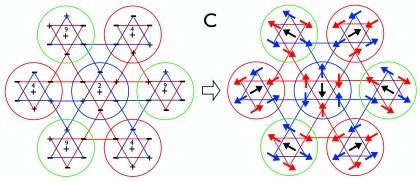}~~~~~
\includegraphics*[width=3in]{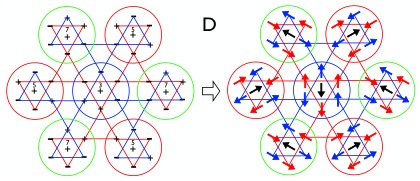}
\caption{ (continued). The way to obtain a collinear bi-pyramid state from a color-sign state. (C) A $\sqrt{3}\times\sqrt{3}$ long range ordered collinear bi-pyramid state from the 2-4-9 sign state. (D) A $\sqrt{3}\times\sqrt{3}$ long range ordered collinear bi-pyramid state from the 3-5-7 sign state.
} \label{S2cont}
\end{figure}

\begin{figure}[H]
\includegraphics*[width=3in]{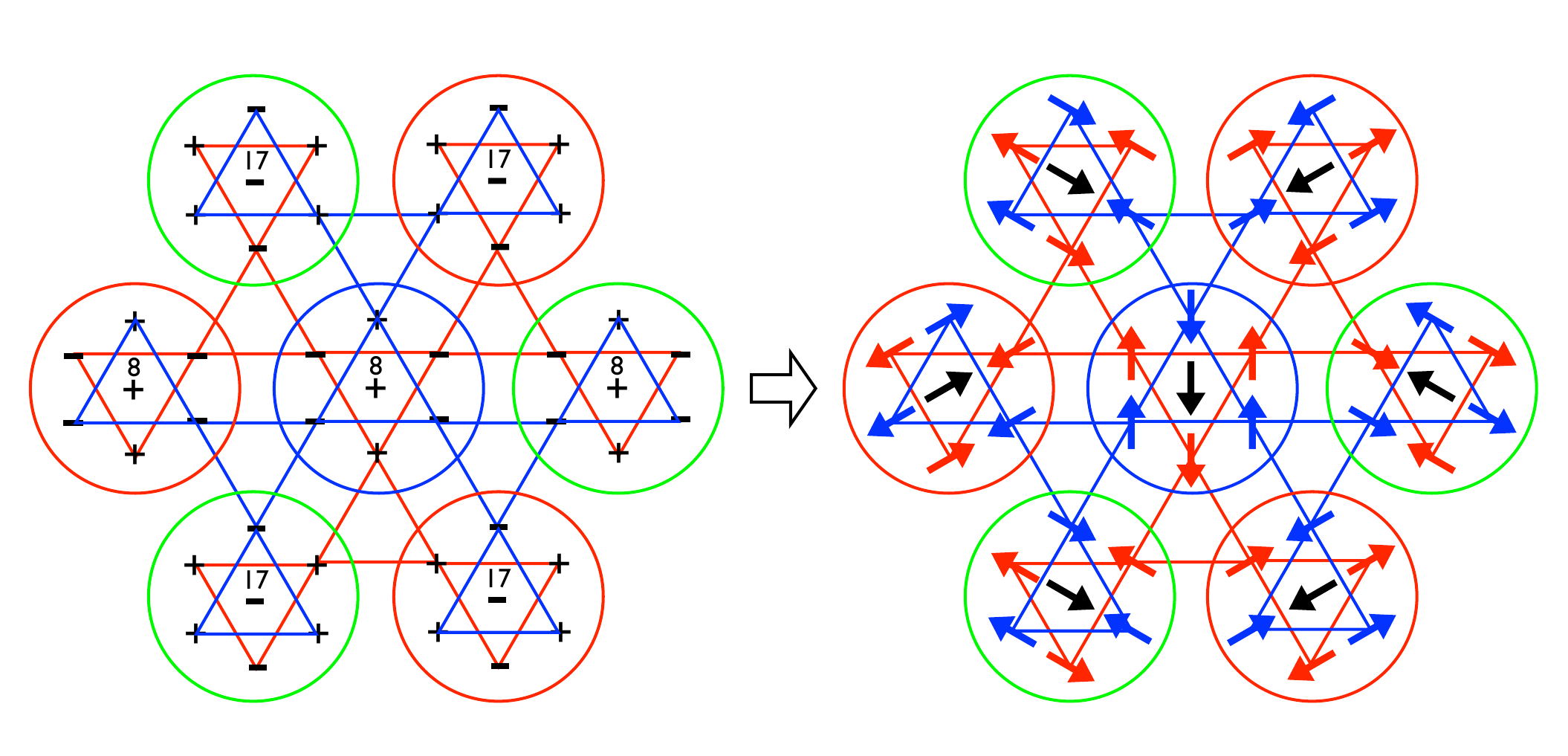}
\caption{An example of a stripe state: alternating rows of 17 and 8 sign states.
} \label{stripes}
\end{figure}

\section{Classical zero energy rotations}
Starting from a classical spin configuration we can rotate simultaneously a subset of the spins, while remaining in the classically degenerate manifold satisfying the zero total spin on the tetrahedra and triangles. Presence of modes involving a finite amount of spins, suggests that at low temperatures the system may be in a liquid state due to quantum tunneling. However, in the bi-pyramid lattice we do not find such modes. The modes involving motion of a minimal amount of spins are most easily seen in the ordered states. Examples for these quasi 1d modes are the ``ladder'' and ``spaghetti'' modes depicted in the figures below.

\begin{figure}[H]
\includegraphics*[width=3in]{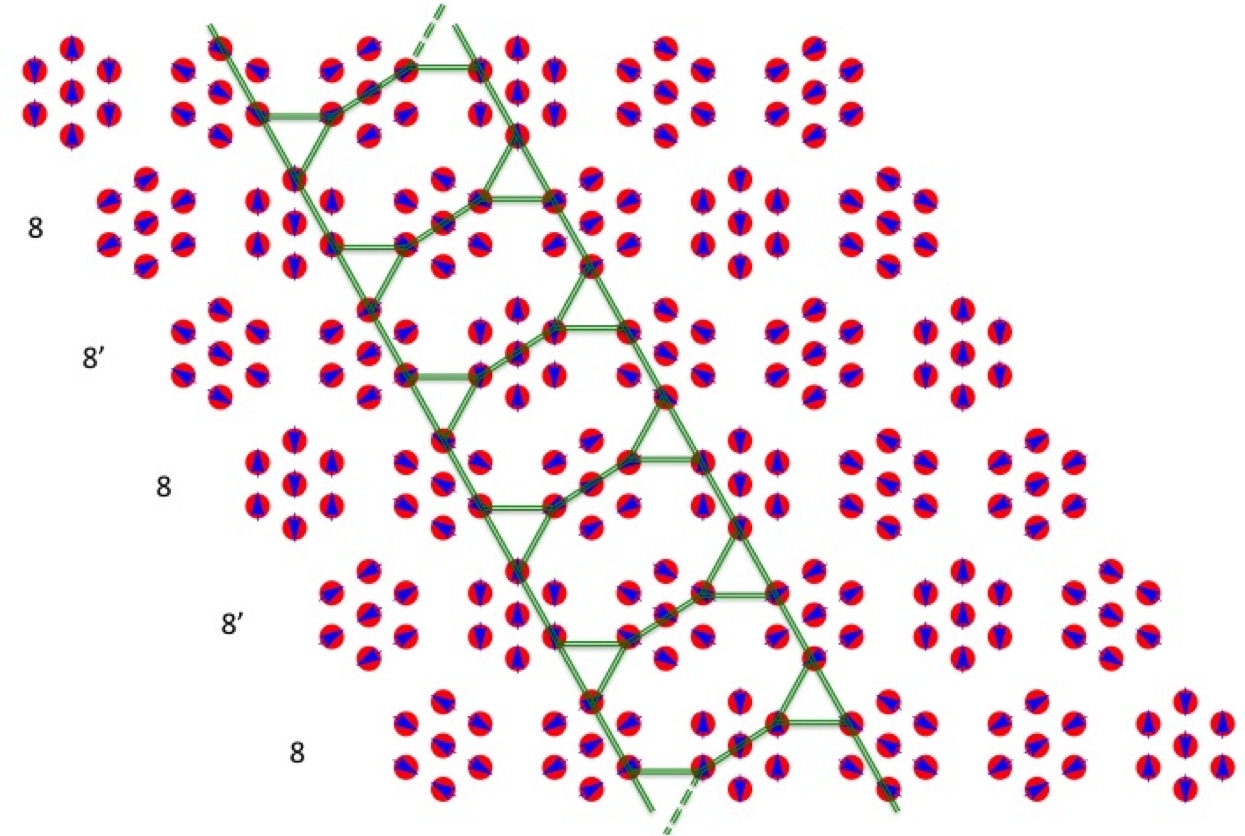}
\caption{An example of a quasi 1d "ladder" mode: the smallest mode (i.e. involving the least number of spins) possible in the ordered stripe phase (here 8' is 8 with a spin flip, equivalent to 17 in Fig. \ref{S1}). The mode involves the simultaneous rotation of the spins marked in green.
} \label{ladderInstripe}
\end{figure}
\begin{figure}[H]
\includegraphics*[width=3in]{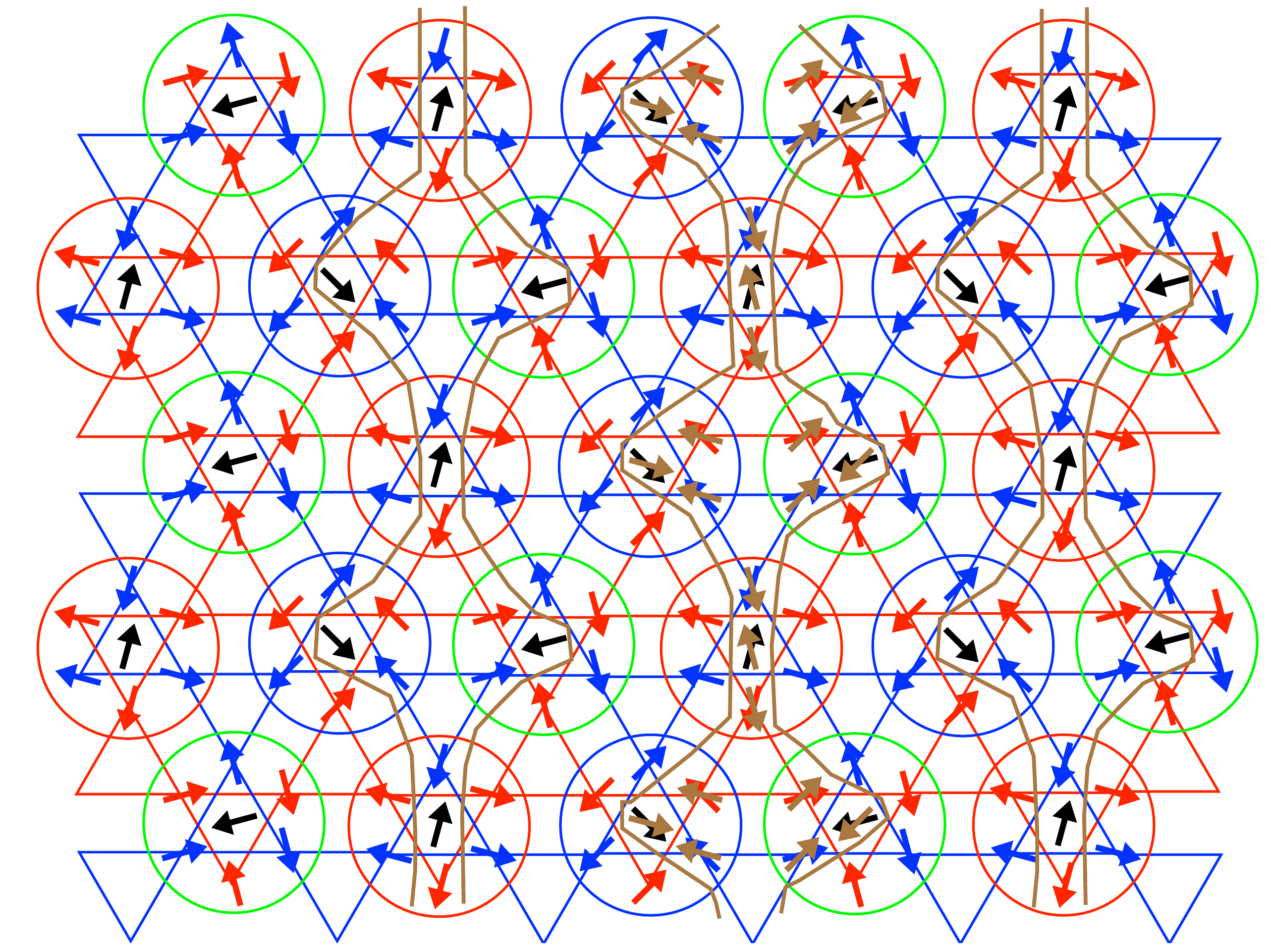}~~~~~~
\includegraphics*[width=3in]{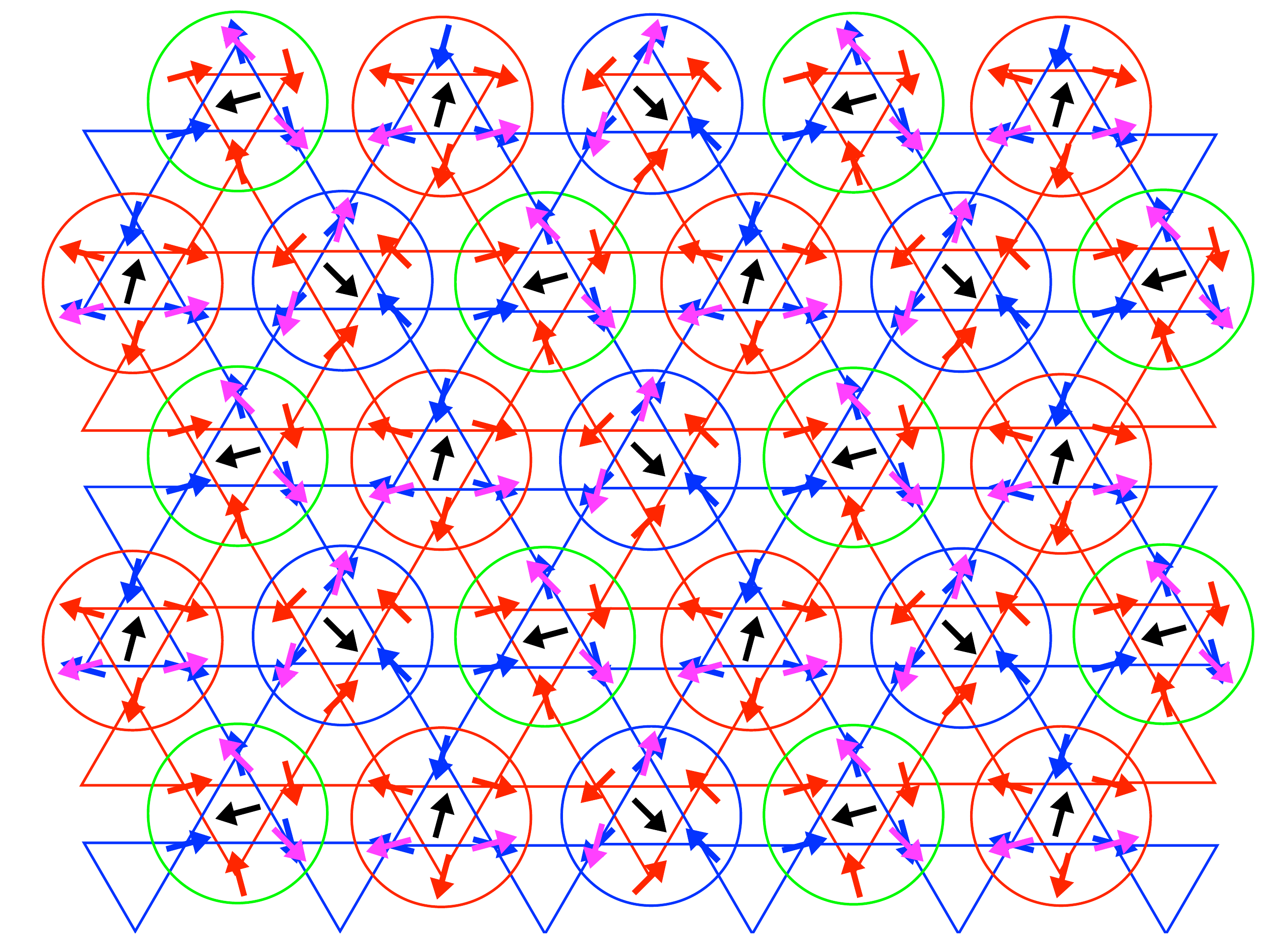}
\caption{A coplanar structure with $\sqrt{3}\times\sqrt{3}$ structure can support partial modes. Left panel: 3 straight ``spaghetti'' modes traversing an ordered state in the vertical direction. Right panel:  simultaneous rotation of $2/3$ of the spins in one kagome layer.
} \label{mode examples}
\end{figure}

\section{Counting states as a tiling problem}
\begin{figure}[H]
\includegraphics*[width=3in]{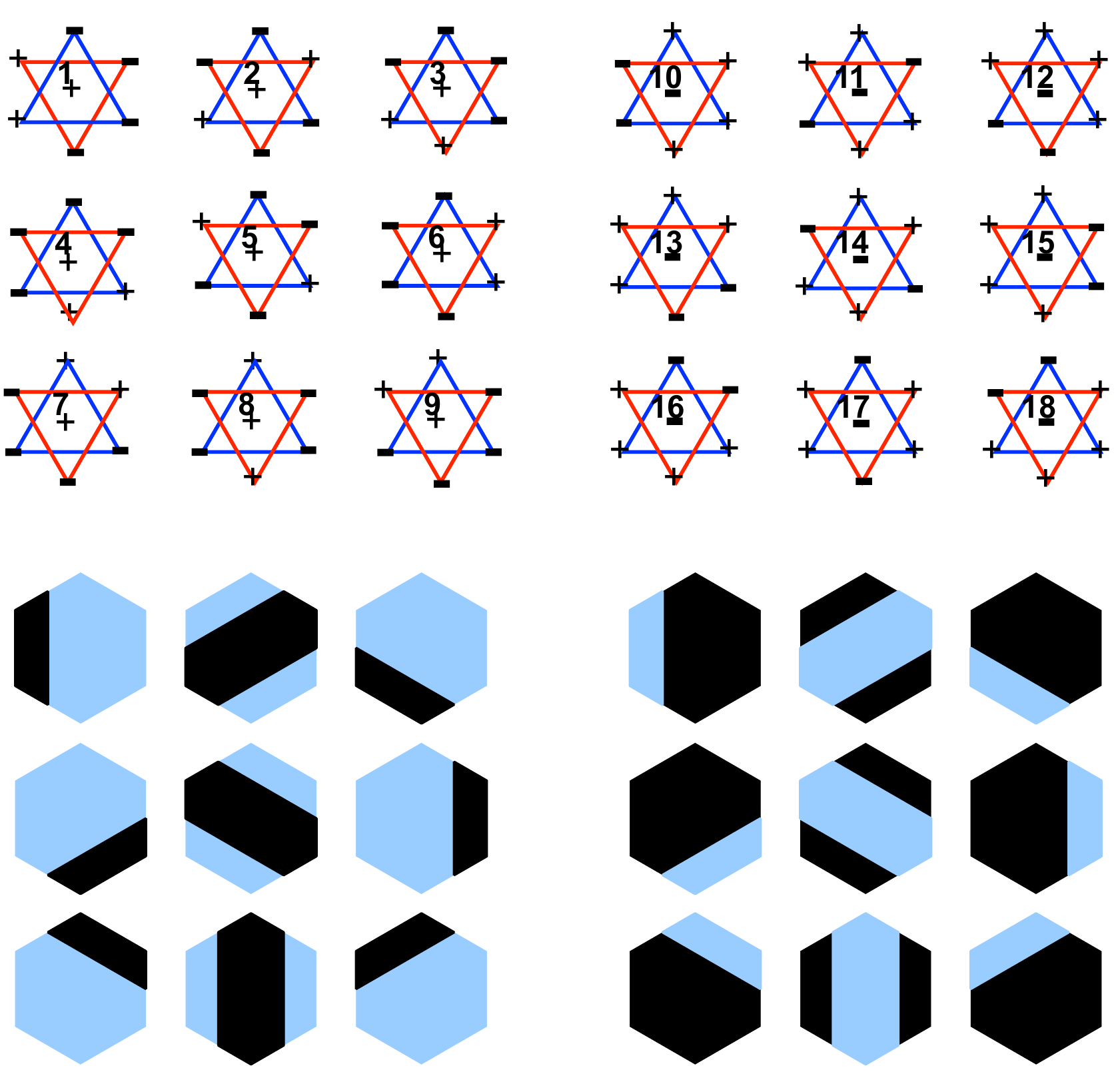}\caption{Tiles for the sign states.
} 
\label{tiles}
\end{figure}

\begin{figure}[H]
\includegraphics*[width=2in]{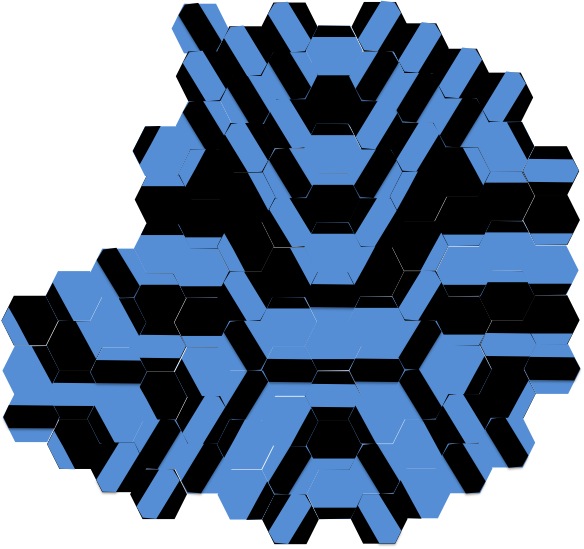}\caption{Tile representation for a random sign state.
} 
\label{typicalnetwork}
\end{figure}

\begin{figure}[H]
\includegraphics*[width=2in]{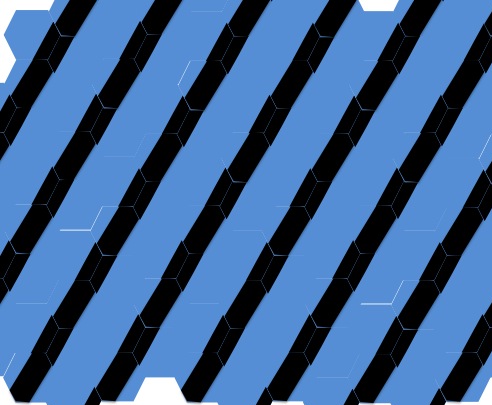}~~~~
\includegraphics*[width=2in]{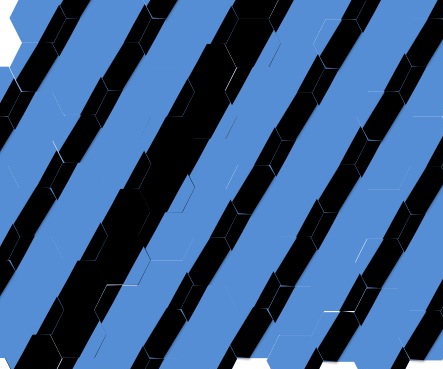}~~~~
\includegraphics*[width=1.8in]{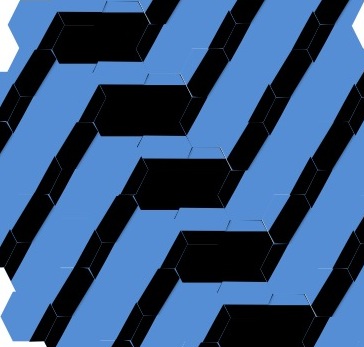}\caption
{Examples of tiling representation of ordered states. Left panel: The $1-6-8$ state represented by tiles. Middle panel: The $1-6-8$ state with a full $\pi$ parallel spaghetti. Right panel: The $1-6-8$ state with a full $\pi$ diagonal spaghetti.
} 
\label{618Pipes}
\end{figure}

In this section we will use an alternative representation of the state in terms of a planar, fully packed pipe network. In this representation, the sign states are represented by the hexagonal tiles depicted in the figure \ref{tiles}.
The rules for tiling are simple: colors must match.
A typical state is shown in the figure
\ref{typicalnetwork}, and we can think of it as a fully packed diagram of pipes. We note that straight segments of pipes may have only two possible thicknesses, which we refer to as thick or thin, and we set as 1 or 2, in appropriate units.
The ordered, 618 state is depicted in \ref{618Pipes}. The same state after putting in a full $\pi$ rotation along a parallel and diagonal spaghetti (see depiction of ``spaghetti'' modes in Fig. \ref{mode examples}), respectively, is shown in the next two figures.

\section{Basic properties of a tiled region}

{\bf Property 1}:
It is impossible to terminate a line. 

\noindent {\it Proof of property 1:} This property can be verified by inspection of possible termination points, and ruling each of them out.

{\bf Property 2}:
There can be no closed loops in a pipe diagram. 

\noindent {\it Proof of property 2:} To prove this, assume the contrary and consider a closed loop of black color, inside there must be loops of smaller and smaller sizes. Since the colors alternate, we must have inside an enclosed simply connected region which is entirely black or entirely blue. as we  don't have a an all blue or all black hexagon, therefore the inner region must be made of lines with termination points. By property 1, this cannot happen.

{\bf Definition:} {\it "Laminar Region"}. 

A Laminar region is a region where lines do not split (i.e. there are no junctions).

Property 3:

The network in a Laminar region is a set of black and blue pipes (not necessarily straight), which are parallel to each other. Moreover, the thickness of each pipes is determined by the thickness at any given point along the pipe.

\begin{figure}[H]
\includegraphics[width=1in]{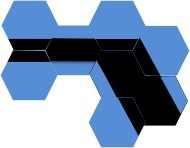}\caption{A direction change without a junction leads always leads to a thickness change.
} 
\label{corner}
\end{figure}

\noindent {\it Proof of property 3:} If no junctions are present, since the lines are fully packed and cannot cross or join, the arrangement is of parallel curves. We now inspect the thickness of the curves. Since the thickness is constant along straight segments, we have to consider elbows (corners). There is only a finite number of possible elbows, which can be checked explicitly. We find that at a corner without a junction, the thickness of a pipe is switched deterministically from thin to thick and vise versa. Thus, given the thickness of the pipe, and the set of points where it has changed direction, the thickness is determined all along the pipe. An example of a corner is in figure \ref{corner}.

A simple example of Laminar regions are the 1-6-8 states, as well as it's variants involving a spaghetti  exhibited in the figures above.

In a Laminar region, property 3 leads to the following immediate estimate of the number of possible states in a region bound by a curve of length $L$. Choosing a reference black curve of length $L$, we can specify the thicknesses of  $N$ of its parallel pipes. The area $A$ of such a configuration is $NL<A<2NL$, and the number of possibilities involved is bounded above by  $3^L2^N$: at each point on the reference curve we can either stay straight or make a $\pi/3$ degree turn left or right, in addition, we have $2^N$ choices of the thicknesses of the parallel curves.
Remark: This, of course is a large over-estimate, since the reference curve cannot self intersect, moreover, it's shape is highly restricted by properties 1,2, which applies to it as well as all the parallel curves. 

Finally, we show that given the configuration of tiles a boundary layer of a thickness 3 hexagons of a laminar region, the state can be completely determined. To see this, note that to establish the state we have to determine the locations of the elbows, as well as the thickness of the lines. The thicknesses are always determined at the boundary. 
For each internal elbow, it's image will appear on the boundary of the region in at least two points  (as it cannot disappear going from one line to the next). If we determine a layer of the boundary thick enough to detect all possible elbows, which can be done with a boundary of thickness 3, we can continue to determine the internal state using the information on the thickness of the lines at the boundary.

\section{Bounding the number of states}
\begin{figure}[H]
\includegraphics*[width=1in]{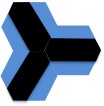}\caption{A junction of narrow black pipes.
} 
\label{NNNjunction}
\end{figure}

By the arguments above, the number of laminar regions scales as the boundary length. To complete our bound on the total number of states, we must include junctions.
There are several kinds of junctions. Most involve a narrow and wide lines meeting. In addition there is a triple narrow junction type (Fig. \ref{NNNjunction}). All junctions of valence larger than 3 may be viewed as joining of two valence 3 junctions.
We now bound the number $N_J$ of possible junctions in the sample.
This can be done by a simple argument: by properties 1,2, the system is a set of lines with no closed cycles, and termination points only on the boundary. Such a system may be viewed as a graph consisting of a disjoint union of trees with leaves on the boundary. Since the number of vertices on a full tree is always smaller then the number of trees we have that:
\begin{eqnarray}
N_j< L
\end{eqnarray}
For each configuration of junctions, we can consider the system consisting of the boundary and a small region around the junctions as a laminar region.  The effective boundary degrees of freedom are less than $\alpha (L+N_j)$, where the constant $\alpha$ describes the "thickening" needed to observe elbows, as before.  Therefore we have at most:
\begin{eqnarray}
N(V,L)<\sum_{N_j=0}^{L}C^V_{N_j} ~c^{(L+N_j)}
\end{eqnarray}
where $C^L_{N_j}$ are binomial coefficients, and $c=18^{\alpha}$. We can easily bound the last equation using
$$
C^n_m <{n^m\over m!}
$$
and get:
\begin{eqnarray}
N(V,L)<c^L\sum_{N_j=0}^{L}{V^{N_j}\over N_j!} <L (cV)^{L}
\end{eqnarray}
for a typical $V\leq \mu L^2$, and we have: 
\begin{eqnarray}
N(V,L) <e^{a_1 L+a_2 L\log L}
\end{eqnarray}

\section{Transfer Matrix Estimates}

In this section we show how locally colinear states may be counted using a transfer matrix method. While we get fairly quickly an upper bound on the number of states, it is extensive in system volume. However, numerics shows that this seems like a large over-estimate: the numerical behavior is quite peculiar and is consistent with a boundary entropy.

Let us consider a description of this problem as follows. 
We consider a square array of the bi-pyramids. It may be viewed as alternating two zig-zag columns which are shifted with respect to each other vertically. 
\begin{figure}[H]
\includegraphics*[width=3in]{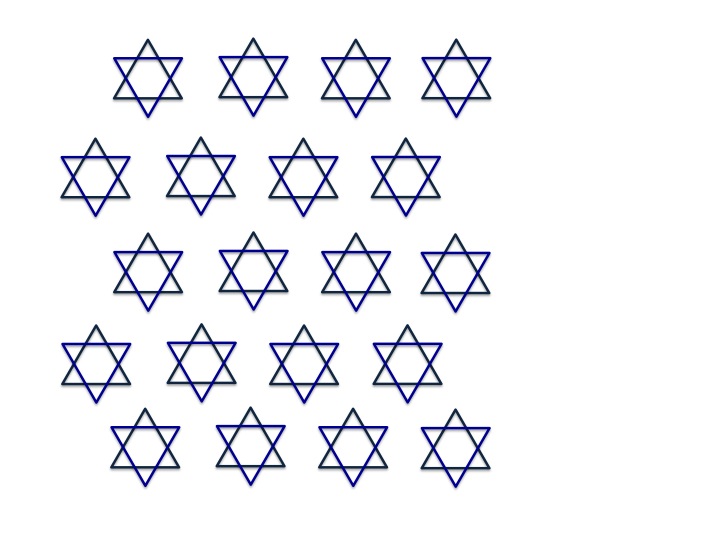}\caption{Square array of bi-pyramids}
\label{bp array}
\end{figure}
We enumerate the possible signs a long the zig-zag columns as follows. We consider adding another column in two steps: we first add sites at even levels and then the sites at odd level as depicted in the fig \ref{twosteps}. 
\begin{figure}[H]
\includegraphics*[width=5in]{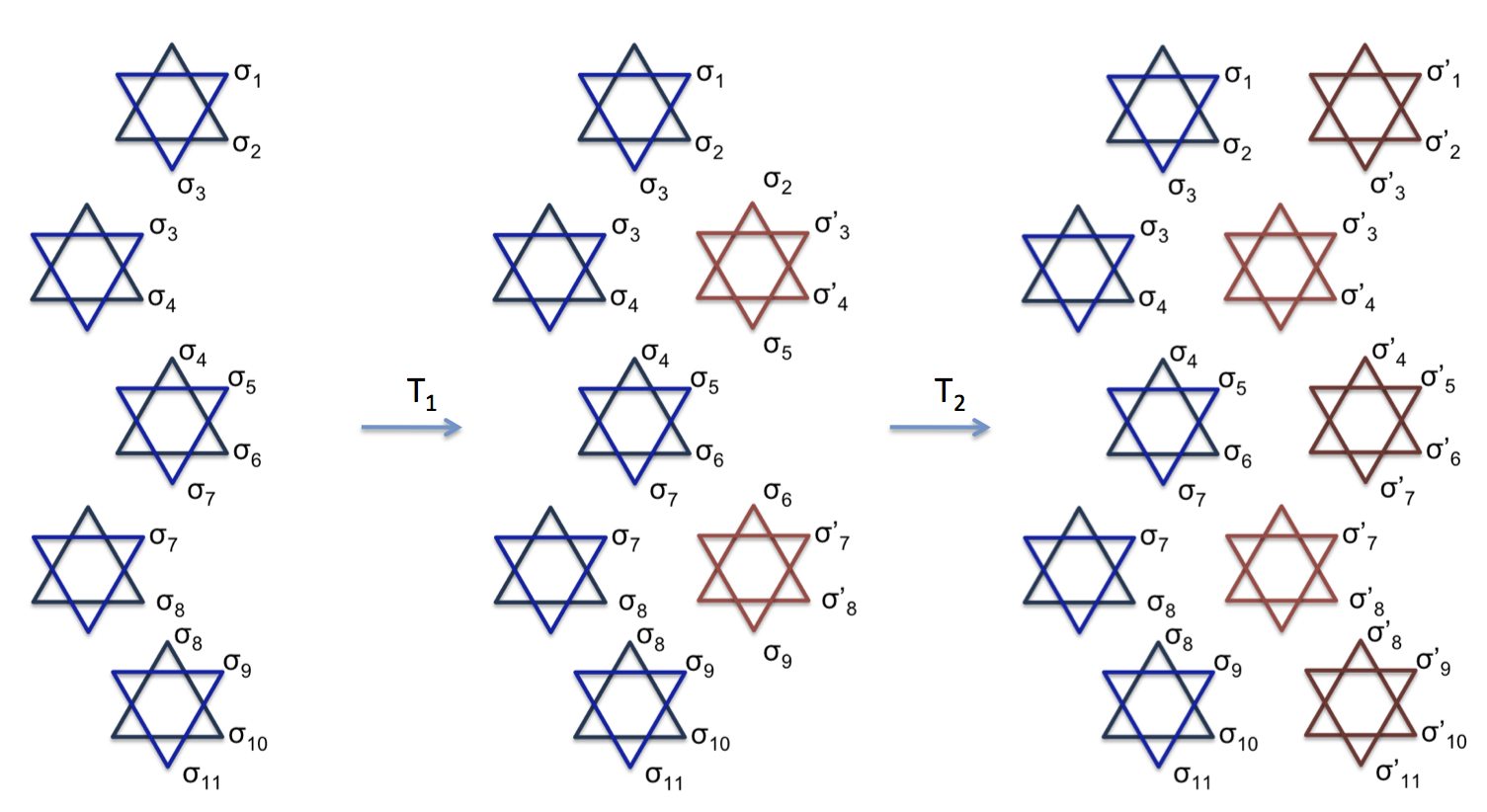}\caption{Construction of the transfer matrix
} 
\label{twosteps}
\end{figure}

\begin{thebibliography}{10}

\bibitem{pauling1935structure}
Pauling L
\newblock (1935) The structure and entropy of ice and of other crystals with
  some randomness of atomic arrangement.
\newblock \emph{Journal of the American Chemical Society} 57:2680--2684.

\bibitem{ramirez1999zero}
Ramirez A, Hayashi A, Cava R, Siddharthan R, Shastry B
\newblock (1999) Zero-point entropy in `spin ice'.
\newblock \emph{Nature} 399:333--335.

\bibitem{bramwell2001spin}
Bramwell ST, Gingras MJ
\newblock (2001) Spin ice state in frustrated magnetic pyrochlore materials.
\newblock \emph{Science} 294:1495--1501.

\bibitem{lee2002emergent}
Lee SH, {et~al.}
\newblock (2002) Emergent excitations in a geometrically frustrated magnet.
\newblock \emph{Nature} 418:856--858.

\bibitem{balents2010spin}
Balents L
\newblock (2010) Spin liquids in frustrated magnets.
\newblock \emph{Nature} 464:199--208.

\bibitem{anderson1973resonating}
Anderson P
\newblock (1973) Resonating valence bonds: A new kind of insulator?
\newblock \emph{Materials Research Bulletin} 8:153--160.

\bibitem{schroers2013bulk}
Schroers J
\newblock (2013) Bulk metallic glasses.
\newblock \emph{Physics Today} 66:32--37.

\bibitem{greer1993confusion}
Greer AL
\newblock (1993) Confusion by design.
\newblock \emph{Nature: International weekly journal of science} 366:303--304.

\bibitem{villain1979insulating}
Villain J
\newblock (1979) Insulating spin glasses.
\newblock \emph{Zeitschrift f{\"u}r Physik B Condensed Matter} 33:31--42.

\bibitem{mydosh1993spin}
Mydosh JA, Barrett TW
\newblock (1993) \emph{Spin glasses: an experimental introduction}
\newblock (Taylor \& Francis London) Vol.{} 125.

\bibitem{bouchaud1993self} Bouchaud JP,  Mézard M
\newblock (1994) \emph{Self induced quenched disorder: a model for the glass transition}
\newblock \emph{J. de Physique I}, 4(8), 1109-1114.

\bibitem{garrahan2010review}  Garrahan JP, Sollich P,  Toninelli C
\newblock (2010) \emph{Kinetically constrained models.} 
\newblock arXiv:1009.6113. Chapter of "Dynamical heterogeneities in glasses, colloids, and granular media", (Oxford University Press).


\bibitem{castelnovo2005quantum} 
Castelnovo C, Chamon C, Mudry C, Pujol P 
\newblock \emph{Quantum three-coloring dimer model and the disruptive effect of quantum glassiness on its line of critical points.}
\newblock \emph{Physical review B} 72(10), 104405 (2005).

\bibitem{chamon2005quantum} 
Chamon C
\newblock \emph{Quantum glassiness in strongly correlated clean systems: an example of topological overprotection.}
\newblock \emph{Physical review letters} 94(4), 040402 (2005).


\bibitem{lee1996spin}
Lee SH, {et~al.}
\newblock (1996) Spin-glass and non--spin-glass features of a geometrically
  frustrated magnet.
\newblock \emph{EPL (Europhysics Letters)} 35:127.

\bibitem{obradors1988magnetic}
Obradors X, {et~al.}
\newblock (1988) Magnetic frustration and lattice dimensionality in
  SrCr8Ga4O19.
\newblock \emph{Solid State Communications} 65:189--192.


\bibitem{ramirez1992elementary}
Ramirez A, Espinosa G, Cooper A
\newblock (1992) Elementary excitations in a diluted antiferromagnetic
  kagom{\'e} lattice.
\newblock \emph{Physical Review B} 45:2505.

\bibitem{lee1996isolated}
Lee SH, {et~al.}
\newblock (1996) Isolated spin pairs and two-dimensional magnetism in
  SrCr9pGa12-9pO19.
\newblock \emph{Physical review letters} 76:4424.

\bibitem{hagemann2001geometric}
Hagemann I, Huang Q, Gao X, Ramirez A, Cava R
\newblock (2001) Geometric magnetic frustration in Ba2Sn2Ga3ZnCr7O22: A
  two-dimensional spinel based kagom{\'e} lattice.
\newblock \emph{Physical review letters} 86:894.


\bibitem{bono2005correlations}
Bono D, Limot L, Mendels P, Collin G, Blanchard N
\newblock (2005) Correlations, spin dynamics, defects: the highly frustrated
  kagom{\'e} bilayer.
\newblock \emph{Low temperature physics} 31:704.

\bibitem{iida2012coexisting}
Iida K, Lee SH, Cheong SW
\newblock (2012) Coexisting order and disorder hidden in a
  quasi-two-dimensional frustrated magnet.
\newblock \emph{Physical Review Letters} 108:217207.

\bibitem{diestel2005graph}
Diestel R
\newblock (2005) Graph theory. 2005.



\bibitem{sen2012vacancy}
Sen A, Damle K, Moessner R
\newblock (2012) Vacancy-induced spin textures and their interactions in a
  classical spin liquid.
\newblock \emph{Physical Review B} 86:205134.


\bibitem{arimori2002ordering}
Arimori T, Kawamura H
\newblock ((2001)) Ordering of the antiferromagnetic heisenberg model on a
  pyrochlore slab.
\newblock \emph{J. Phys. Soc. Jpn.} 70:3695--3707.

\bibitem{halperin1977hydrodynamic}
Halperin BI, Saslow WM
\newblock (1977) \emph{Hydrodynamic theory of spin waves in spin glasses and other systems with noncollinear spin orientations.}
\newblock \emph{ Phys. Rev. B} 16(5) 2154-2162.

\bibitem{anderson1972anomalous} 
Anderson PW, Halperin BI, Varma CM
\newblock (1972) \emph{Anomalous low-temperature thermal properties of glasses and spin glasses.}
\newblock \emph{ Phil. Magazine}, 25(1), 1-9.

\bibitem{podolsky2009halperin}
Podolsky D,  Kim YB
\newblock  (2009) \emph{Halperin-Saslow modes as the origin of the low-temperature anomaly in NiGa2S4.} 
\newblock \emph{Phys. Rev. B}, 79(14), 140402—1-4.

\bibitem{ramirez2000entropy}
Ramirez AP, Hessen B, Winklemann M
\newblock (2000) \emph{Entropy balance and evidence for local spin singlets in a kagome-like magnet.}
\newblock \emph{ Phys. Rev. Lett.} 84, 2957-2960.

\bibitem{baumer2013glass}
Baumer R, Demkowicz M
\newblock (2013) \emph{Glass transition by gelation in a phase separating binary
  alloy.}
\newblock \emph{Physical Review Letters} 110:145502.

\bibitem{bryngelson1987spin}
Bryngelson JD, Wolynes PG
\newblock (1987) \emph{Spin glasses and the statistical mechanics of protein folding.}
\newblock \emph{Proceedings of the National Academy of Sciences} 84:7524--7528.


\bibitem{Blunt2008Random}
Blunt MO et al. 
\newblock (2008)  \emph{Random tiling and topological defects in a two-dimensional molecular network.}\newblock \emph{Science} 322:1077--1081.

\bibitem{Garrahan2009Molecular}
Garrahan JP, Stannard A, Blunt MO, Beton PH
\newblock (2009) \emph{Molecular random tilings as glasses.}
\newblock \emph{ Proc. Natl. Acad. Sci.} USA 106:15209--15213.

\bibitem{Jack2005Caging} 
Jack RL, Garrahan JP
\newblock (2005) \emph{Caging and mosaic length scales in plaquette spin models of glasses.}
\newblock \emph{Journal of Chemical Physics} 123, 164508.

\bibitem{Tchernyshyov2003Bond} 
Tchernyshyov O, Starykh OA, Moessner R, Abanov AG
\newblock (2003) \emph{Bond order from disorder in the planar pyrochlore magnet.}
\newblock \emph{Physical Review B} 68(14), 144422.

\bibitem{Tchernyshyov2004Valence} 
Tchernyshyov O, Yao  H, Moessner R 
\newblock (2004) \emph{Valence-bond crystal in a {111} slice of the pyrochlore antiferromagnet.}
\newblock \emph{Physical Review B} 69:212402—1-4.

\bibitem{Damle2006Spin} 
Damle K, Senthil T
\newblock (2006) \emph{Spin nematics and magnetization plateau transition in anisotropic kagome magnets.} \newblock \emph{Physical Review Letters}, 97(6), 067202.

\bibitem{Xu2007Global} 
Xu C, Moore JE
\newblock (2007) \emph{Global phase diagram for the spin-1 antiferromagnet with uniaxial anisotropy on the kagome lattice.}
\newblock \emph{Physical Review B} 76, 104427.

\bibitem{bekenstein2003information}
Bekenstein JD
\newblock (2003) Information in the holographic universe.
\newblock \emph{Scientific American} 289:58--65.

\bibitem{huijse2008charge}
Huijse L, Halverson J, Fendley P, Schoutens K
\newblock (2008) Charge frustration and quantum criticality for strongly
  correlated fermions.
\newblock \emph{Physical review letters} 101:146406.

\bibitem{bombelli1986quantum}
Bombelli L, Koul R, Lee J, Sorkin R
\newblock (1986) {Quantum source of entropy for black holes}.
\newblock \emph{Physical Review D} 34:373--383.

\bibitem{wolf2006violation}
Wolf M
\newblock (2006) {Violation of the entropic area law for fermions}.
\newblock \emph{Physical review letters} 96:10404.

\bibitem{gioev2006entanglement}
Gioev D, Klich I
\newblock (2006) {Entanglement entropy of fermions in any dimension and the
  Widom conjecture}.
\newblock \emph{Phys. Rev. Lett.} 96:100503.

\end{thebibliography}
In this procedure all the sites in the first stage can be added independently of each other, and after it is complete the second stage shares this property.
We are now left with the task of transferring this into a formula.
For a bi-pyramids in the first move, we note that the signs  ${\sigma}_1,{\sigma}_2,{\sigma}_5,{\sigma}_6..$ and in general ${\sigma}_{4n+1},{\sigma}_{4n+2}$, where $n$ is an integer, remain unchanged. The sites which might change after the move are of the form ${\sigma}_{4n-1},{\sigma}_{4n}$. Each such pair only depends on the states of ${\sigma}_{4n-2},..{\sigma}_{4n+1}$. Let us denote by $M$ the number of possible sign states with the signs on the left ${\sigma}_{4n-2},..{\sigma}_{4n+1}$ and sites on the right: ${\sigma'}_{4n-1},{\sigma'}_{4n}$.

We can view this as a linear transformation ${\cal M}$ on ${\mathbb Z}^{\otimes 4}$
\begin{eqnarray} {\cal M}|{\sigma}_{4n-2},{\sigma}_{4n-1},{\sigma}_{4n},{\sigma}_{4n+1}\rangle=M |{\sigma}_{4n-2},{\sigma'}_{4n-1},{\sigma'}_{4n},{\sigma}_{4n+1}\rangle\end{eqnarray} 
For a column of  $N_y$ bi-pyramids, there are $2N_y+1$ signs on the border which participate in the counting.
In the first stage we can combine all the $\cal M$ moves to
\begin{eqnarray}
T_1=\mathbb{I}\otimes \cal M\otimes \cal M...\otimes \cal M\otimes \mathbb{I}\otimes \mathbb{I}
\end{eqnarray}
Next we note that the transformation governing the added bi-pyramids in step 2, is described in the same, albeit shifted by two sites. In addition, it involves adding boundary bi-pyramids, which require special counting. We can summarize this as:
\begin{eqnarray}
T_2= {\cal M}_3\otimes {\cal M}...\otimes {\cal M}\otimes {\cal M'}_3
\end{eqnarray}
For $N_x$ columns, the number of states may now be computed as 
\begin{eqnarray}
N=\langle R| (T_2T_1)^{N_x}|L\rangle
\end{eqnarray}
where $|L\rangle,|R\rangle$ specify boundary conditions on the left and on the right.

Now, the number of states in a large region, $N_x\rightarrow\infty$ scales as $\lambda(N_y)_{max}^{N_x}$, where $\lambda(N_y)_{max}$ is the largest eigenvalue of $T_2T_1$. Next we consider the eigenvalues of $T_2$ and $T_1$ separately.

Assuming $N_y$ is large, these are determined by $\cal M$. The matrix $\cal M$ is special, and it's eigenvalues can be determined analytically. 
To do so we first determine the invariant subspaces of this matrix. 
$\cal M$ is a $16\times 16$ matrix, acting on 4 ising spins. It turns out more convenient to write it using two double spins, in basis 4 by assigning: 
\begin{eqnarray}
\left\{\sigma _1,\sigma _2,\sigma _3,\sigma _4\right\}\to \left\{\sigma _1+2 \sigma _2+1,\sigma _3+2 \sigma _4+1\right\}
\end{eqnarray}
We now find the cycles of the matrix:
\begin{eqnarray}
\left|
\begin{array}{c|c|c}\hline
 \text{Involved states} & \text{Moves} & \text{eigenvalues} \\ \hline
 32;11 & \{3,2\}\rightarrow\{3,2\} ; \{3,2\}\rightarrow\{1,1\} ; \{1,1\}\rightarrow\{3,2\} & \frac{1}{2} \left(1\pm\sqrt{5}\right) \\
 21;41;42 & \{2,1\}\rightarrow\{4,1\} ; \{4,1\}\rightarrow\{2,1\};
\{4,2\}\rightarrow \{4,1\};\{4,1\}\rightarrow \{4,2\} & \pm \sqrt{2},0 \\
 31;12 &  \{3,1\}\rightarrow\{1,2\};\{1,2\}\rightarrow\{3,1\} & \pm 1 \\
 34;14;13 &  \{3,4\}\rightarrow\{1,4\};\{1,4\}\rightarrow\{3,4\};
\{1,4\}\rightarrow\{1,3\};\{1,3\}\rightarrow\{1,4\} & \pm \sqrt{2},0 \\
 23;44 & \{2,3\}\rightarrow\{2,3\};\{2,3\}\rightarrow\{4,4\};\{4,4\}\rightarrow\{2,3\} & \frac{1}{2} \left(1\pm\sqrt{5}\right) \\
 24;43 & \{2,4\}\rightarrow\{4,3\};\{4,3\}\rightarrow\{2,4\} & \pm 1 \\ \hline
\end{array}
\right|
\end{eqnarray}

Interestingly, the 6 largest eigenvalues are $\phi,\sqrt{2},1$, each doubly degenerate. Here the largest eigenvalue is $\phi={1+\sqrt{5}\over 2}$ is the golden ratio.
By the norm inequality:
$||T_1T_2||\leq ||T_1|| ||T_2||$, we immediately conclude that the largest eigenvalue of $T_1$ and of $T_2$ scale as $\phi^{N_y \over 2}$. From this we have a rough estimate that the number of states scales at most as $\phi^{N_x N_y}$.

\subsection{Numerical behavior of the eigenvalues of the transfer matrices}

The largest eigenvalues of transfer matrices can be computed numerically up to several rows. Here are the two largest eigenvalues of the transfer matrix $T_1T_2$ up to 11 rows:
\begin{eqnarray}
T_1T_2=(M_4\otimes..\otimes M_4)(M_{3up}\otimes M_4\otimes..\otimes M_4\otimes M_{3down})
\end{eqnarray}

\begin{eqnarray}
\left|
\begin{array}{c|c|c}\hline
 L_y & \lambda _1 & \lambda _2 \\ \hline
 1 & \frac{1}{2} \left(5+\sqrt{17}\right)\sim 4.56 & -2 \\
 3 & 6.7966 & 6.7966 \\
 5 & 6.26138 & 6.26138 \\
 7 & 6.04937 & 6.04937 \\
 9 & 5.99248 & 5.99248 \\
 11 & 5.95587 & -2.62806+4.77494 i \\ \hline
\end{array}
\right|
\end{eqnarray}
It is very clear that, at least up to 11 rows, the largest eigenvalue goes down. This means, that for a long enough strip, the number of states of 11 rows, will be smaller than, say the number of states of 3 rows.
This reflects the highly constrained nature of the system: for certain sign configuration of the 3 rows, there can be no way to continue adding rows in a consistent way up to 11 rows. 

The decreasing nature the eigenvalues give us strong evidence that, in fact, the number of states should scale as the boundary length, without an extra $log$.

\section{Coplanar bi-pyramid states}
\begin{figure}[H]
\includegraphics*[width=3.7in]{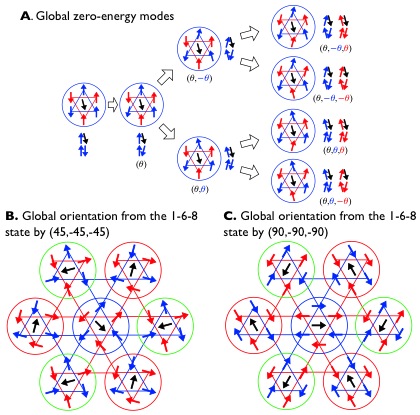}
\caption{The way to obtain coplanar bi-pyramid states from collinear bi-pyramid states. (A) For one bi-pyramid, there are four different ways of generating coplanar state from a collinear state when all spins rotate. (B) and (C) show two examples obtained by such rotations starting from the 1-6-8 sign state. 
} 
\label{S3}
\end{figure}

For one bi-pyramid, a coplanar state can be generated from a collinear state by rotating the 7 spins in several collective ways, four of which with a set of parameters $(\theta,\pm\theta,\pm\theta)$ are shown in Fig. \ref{S3}A. For the entire triangular lattice of bi-pyramids, because of the color and ferro-sign bond constraints, the three angles with the same magnitude are sufficient to generate a long range ordered coplanar state. Fig. \ref{S3}B shows one orthogonal bi-pyramid spin state, as an example, generated from the 1-6-8 sign state by rotating the spins with (45,-45,-45). If the spins are rotated with (90,-90,-90), then the bi-pyramids become collinear again, as shown as an example in Fig. \ref{S3}C that corresponds to the 7-3-5 sign state. The collinear spin states and their resulting coplanar states are continuously connected with each other in the spin-energy phase space. The connections among the sign states are listed in the table below.

\begin{tabular}{ l | c | r }
\hline\hline
Initial sign state	&Rotation angles &	Final sign state\\ \hline
(1-6-8) &	(90,-90,-90)&	(3-5-7)\\
(1-6-8) &	(90,-90,90)&	(1-6-8)\\
(1-6-8) &	(90,90,90)&	(4-9-2)\\
(1-6-8) &	(90,90,-90)&	(1-6-8)\\
(2-4-9)&	(90,-90,-90)&	(1-6-8)\\
(2-4-9)&	(90,-90,90)&	(2-4-9)\\
(2-4-9)&	(90,90,90)&	(3-5-7)\\
(2-4-9)&	(90,90,-90)&	(2-4-9)\\
(3-5-7)&	(90,-90,-90)&	(3-5-7)\\
(3-5-7)&	(90,-90,90)&	(2-4-9)\\
(3-5-7)&	(90,90,90)&	(3-5-7)\\
(3-5-7)&	(90,90,-90)&	(1-6-8)\\ \hline
\end{tabular}

The initial and final long-range-ordered sign states connected by the global spin zero-energy excitations illustrated in Fig. \ref{S3}.\section{Spin Wave Calculations}

In order to explore the energy landscape generated by quantum fluctuations, we next concentrated on evaluating the spin wave energy for non uniform systems. 
The calculations have been done in the framework of the Holstein-Primakoff representation. We consider the Hamiltonian:
\begin{eqnarray}
H=\sum_{\langle i j\rangle} J \overrightarrow{S}_i\cdot \overrightarrow{S}_j .
\end{eqnarray}
First we pick a classical spin configuration which is a local energy minimum. 
In the next step we take each classical spin direction, and replace by an operator as:
\begin{eqnarray}
\overrightarrow{S}_i=\hbar (S-a_i^{\dag}a_i)\hat{S}_{class, i}+\hbar \sqrt{2S-a_i^{\dag}a_i}~a_i~\hat{e}_{1}+\hbar a_i^{\dag}\sqrt{2S-a_i^{\dag}a_i}~\hat{e}_{2}
\end{eqnarray}
where $a_i^{\dag},a_i$ are boson creation/annihilation operators, and $\hat{e}_{1},\hat{e}_{2}$ are any couple of unit vectors which combine into an orthogonal frame with 
the classical direction $\hat{S}_{class, i}$. At this point, in order to get a tractable theory, the square roots are expanded to lowest order in $1/S$. Since it is assumed that the spin is at a classical minimum, this procedure yields a quadratic hamiltonian in the bosonic operators.

Uniform (long range ordered states) are studied, as usual, by rewriting the Hamiltonian in momentum space. The number of degrees of freedom, is then determined simply by the unit size. We then compute the zero point energy of the resulting $k$ modes, and sum them over the Brillouin zone.

An example of such a calculation is exhibited in the figure, where different long range ordered states obtained from $618$ are compared.
\begin{figure}[H]
\includegraphics*[width=4.6in]{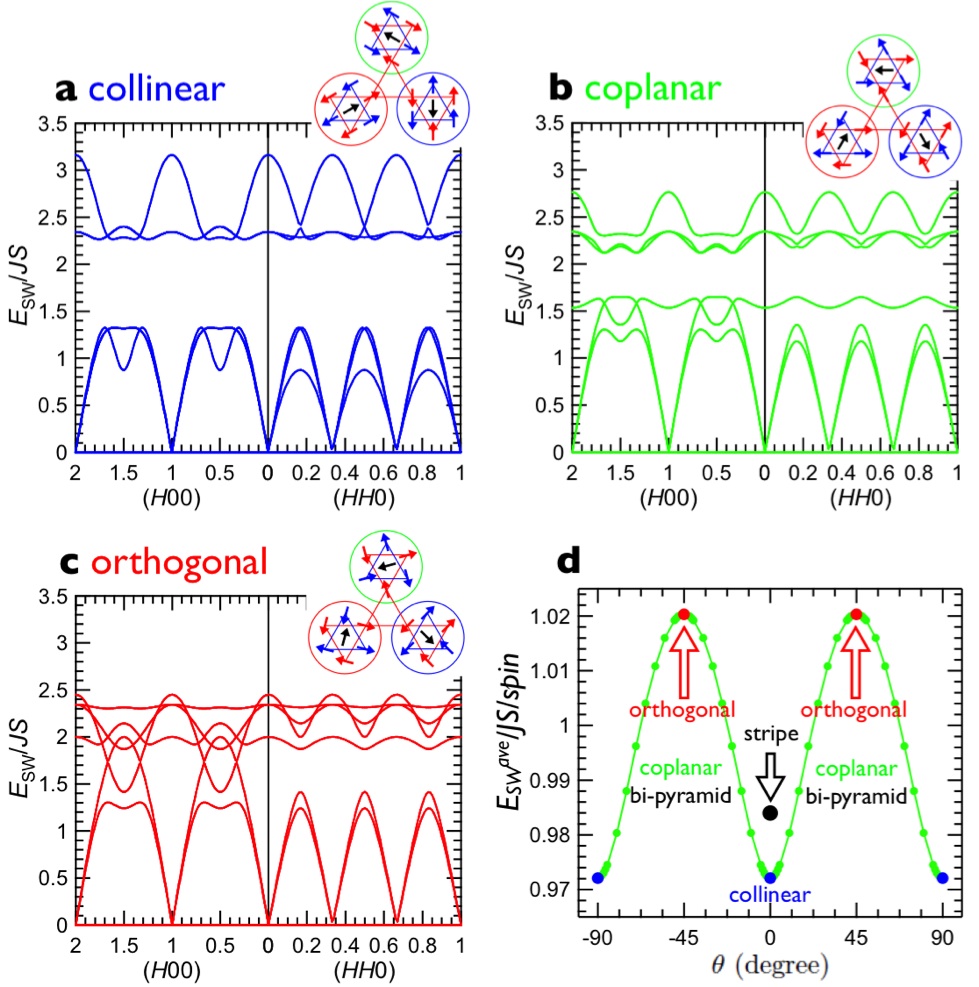}\caption{ (a),(b),(c)  Spin wave dispersions for long range ordered states obtained from the $618$ state by a global reorientation as described in Fig \ref{S3}. (d) Total spin wave energy as function of rotation angle. Maximum is obtained for orthogonal configurations and minimum for collinear. We also show the energy obtained for the stripe state consisting of alternating rows of 8 and 17 type of sign states (see Fig. \ref{stripes} and \ref{ladderInstripe})}
\end{figure}

To consider states which are not translationally invariant requires a real space treatment.  
To diagonalize the Hamiltonian, on a lattice with $N=L_xL_y$ bi-pyramids, we get a   $7L_xLy$ quadratic boson hamiltonian.

To compute the state of the system, we wrote a procedure affecting the symplectic transformations needed to diagonalize the Hamiltonian numerically. We then computed the spin wave energy for the Hamiltonian for various system sizes (as large as $24\times 24$ bi-pyramids, involving a total of 4032 spins).

\begin{figure}[H]
\includegraphics*[width=2.5in]{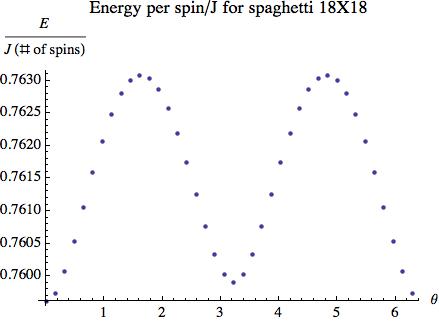}
\includegraphics*[width=4.5in]{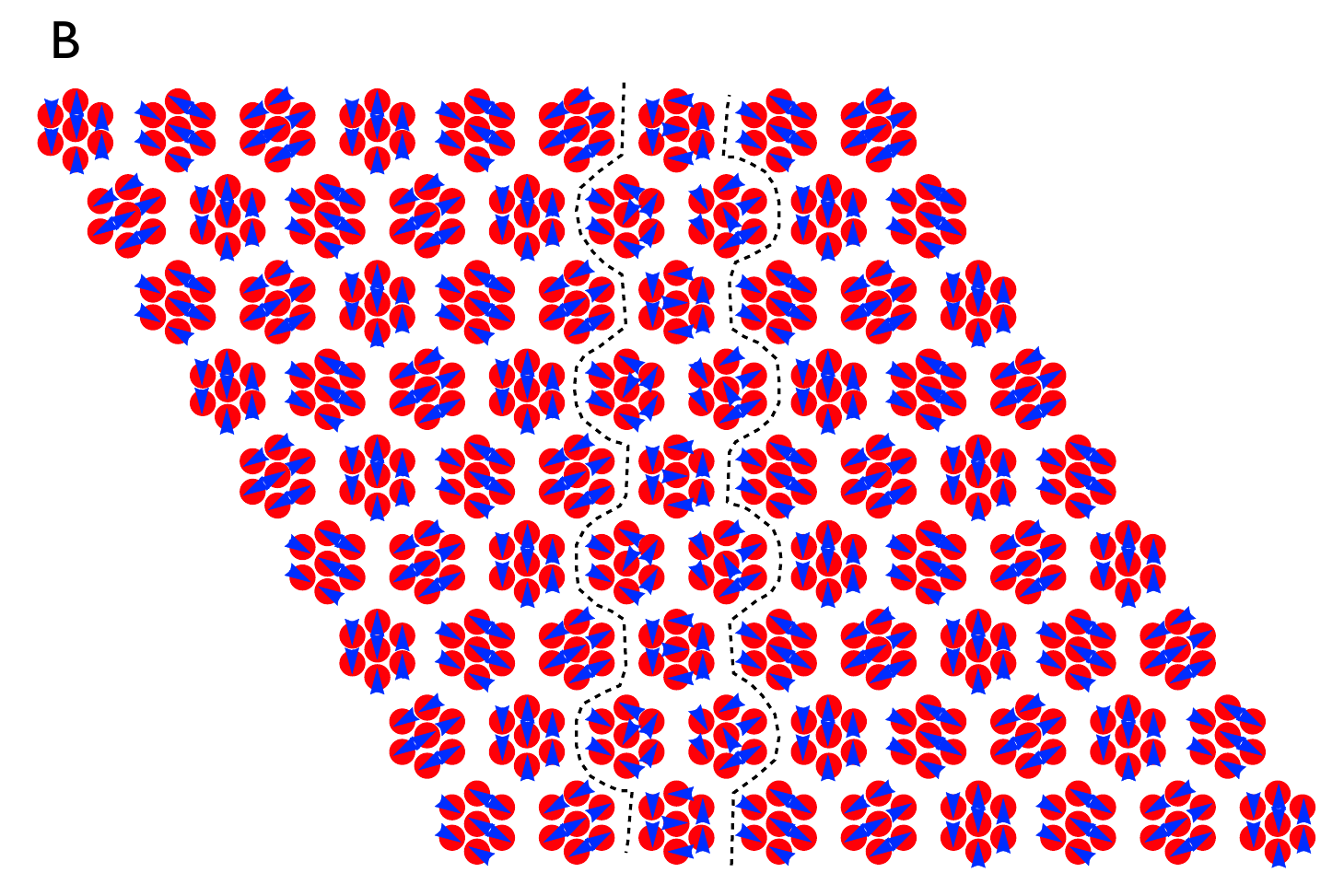}\caption{Spaghetti mode inserted through a 6-1-8 state, as function of angle. The energy is divided by the total volume in this graph. The mode is visible in 
this $9\times 9$ array as a distortion line in the middle of the sample.
} 
\label{S5}
\end{figure}

Figure \ref{S5} shows the insertion of a spaghetti mode (see depiction in Fig. \ref{mode examples}) into an ordered state, and the system energy as function of the angle of rotation of the spaghetti.

\end{document}